\documentstyle[preprint,aps]{revtex}
\begin{document}
\tightenlines
\draft
\preprint{LMU-TPW-97/19}
\title{The string scale and the Planck scale}
\author{Rainer Dick}
\address{Sektion Physik der Ludwig--Maximilians--Universit\"at 
M\"unchen\\
Theresienstr.\ 37, 80333 M\"unchen, Germany}
\maketitle
\begin{abstract}
A particle spectrum below the string scale in accordance with
predictions from heterotic string theory
yields
a Planck mass $m_{Pl}=(8\pi G_N)^{-1/2}$
which exceeds the string scale by a factor $\simeq 61.9$.
A Planck mass $m_{Pl}=2.43\times 10^{18}$ GeV then corresponds 
to a string scale
 $m_s=3.9\times 10^{16}$ GeV. Such a low value for the string scale 
in turn implies
that the relative strength of graviton 
and vector exchange in the string/M-theory phase exceeds the 
corresponding ratio in the
low energy field theory.
\end{abstract}
\pacs{PACS numbers: 11.25.-w, 04.60.-m}

Among all contemporary attempts to extend our knowledge all the 
way up
to scales where quantum gravitational effects are expected to play 
a role string
theory provides the most advanced and best developed 
theoretical framework.
In recent years the field was turmoiled by the emergence of 
M-theory as
a non-perturbative formulation eventually unifying all known consistent
string theories \cite{M1} 
(see also \cite{pre} for early work on the role of eleven dimensions
in string theory and the appearance of matrix models in membrane dynamics), 
and one may ask whether string theory can be expected to
provide any answers to string scale questions in its present form. However,
the subject develops at a fast pace from a phase of existence
proofs into a phase where dynamical issues can be addressed \cite{M1,Mg,M2}
although some conjectural dualities between different sectors of the theory
still await further
confirmation. For these
reasons it might be more appropriate to denote the transition scale 
between energies where low energy field theory formulations
apply and M-theory as an M-scale, but here I will still use the notion of
string scale.
I should also point out that in the present letter I'm only considering
the string scale from the low energy perspective, asking at 
which temperature
we might expect resolution of extended objects. Therefore all assertions
in the present paper only rely on the assumption that all low energy
 field theories arising from M-theory are supergravity
and those supergravity-Yang-Mills-systems inherited from the 
known consistent
superstring theories. If another field theory limit would emerge with a larger
low mass field content than the heterotic string this would increase 
the ratio $m_{Pl}/m_s$ above the value calculated here. 

Since we will consider
the string scale from a low energy perspective no genuine
M-theoretic tools will be needed in the present paper and 
I would like to refer
the reader
to some very useful recent reviews on duality and 
M-theory for the high energy
 formulation of the theory \cite{rev}. From a phenomenological 
point of view it still seems 
reasonable to assume
that the transition to low energy field theories proceeds through
the heterotic string spectrum \cite{GHMR1,GHMR2,revhet}, 
and here I would like to point
out that this gives an interesting new estimate on the relation
between the string scale and the Planck scale from a thermodynamical
point of view. It also turns out that simple
thermodynamical properties of the transition from M-theory to a field theory
description favor the heterotic phase.

The connection between the string scale, the Planck scale and the GUT scale
is a long standing issue in string theory. The role of a separate
GUT scale always remained somewhat obscure
and was rather puzzling from a string theory perspective.
In the present letter I would like to point out 
that the large number of degrees of freedom
of the heterotic string implies a remarkable coincidence
of the string scale with the usual estimate on the GUT scale from 
supersymmetric $\beta$
 functions \cite{GUT}. However, I would also like to stress that Witten recently
proposed another intriguing solution to the GUT scale puzzle:
In \cite{Mg} Witten points out that identification of the GUT scale with a 
Kaluza--Klein scale describing compactification from eleven to five dimensions
and subsequent compactification from five to four dimensions at a scale 
somewhat below $m_{GUT}$
yields a ratio for $m_{Pl}/m_{GUT}$
in agreement with the data if the 
eleven-dimensional Planck mass $m_{Pl,11}$
is slightly above the GUT 
scale ($m_{Pl,11}\simeq 2m_{GUT}\simeq 14m_{5\to4}$
would work, e.g.).
In this scenario the GUT scale looks like a Kaluza--Klein scale rather
than a string scale,
but this is only an artifact of the long wavelength approximation. 
In the present approach, on the other hand, field 
theoretical compactifications
play no role, and
it will be very interesting to further elucidate the relationship between 
the field theoretic
investigations in \cite{Mg} and the present considerations based 
on heterotic
string thermodynamics.

Heterotic string theory made its first
contact with M-theory through the identification with 
M-theory compactified
on an interval \cite{HW}.
This observation established a link to string inspired phenomenology
and boosted conjectures that M-theory
may really provide a unifying framework for all consistent superstring theories.
Among all known field theory formulations of string theory or M-theory the 
heterotic $E_8\times E_8$
theory is unique for its interesting gauge sector and its large number of 
massless degrees
of freedom.
In the particle sector
the heterotic string  predicts 4032 bosonic massless degrees 
of freedom and the same number of fermionic
degrees of freedom, and the effective number of relativistic
degrees of freedom
amounts to $g_{\ast}(T_s)=7560$. This exceeds the effective number
of relativistic type {\sl II} or type {\sl I}
degrees of freedom $g_{\ast II}(T_s)=2g_{\ast I}(T_s)=240$ 
considerably, 
and as a consequence the heterotic string predicts the lowest
value of the string scale for a given value of the Planck scale:

The early phase of the universe below the string scale is radiation dominated
and evolving back present energy densities we know that curvature
contributions are negligible during radiation dominance. 
The energy density during this early phase is then
\begin{equation}\label{eq1}
\varrho=\frac{3m_{Pl}^2}{4t^2}=\frac{\,\pi^2}{30}g_{\ast}(T)T^4
\end{equation}
where $m_{Pl}=(8\pi G_N)^{-1/2}$ is the reduced Planck mass and $t$ 
is the time parameter in the Friedmann--Robertson--Walker
line element. 

Eq.\ (\ref{eq1}) tells us
how the string scale $m_s=1/t_s$ 
relates to the string temperature
and $m_{Pl}$: 
\begin{equation}\label{eq2}
m_{Pl}m_{s}=57.6\,T_s^2
\end{equation}
and these simple thermodynamical properties of the 
heterotic string provide a strong hint for coincidence of the GUT scale
with the string scale: We expect that the extension of strings
becomes visible when the energy per degree 
of freedom
is powerful enough to resolve the string length $t_s$.
In the relativistic domain the average energy $\epsilon$ per 
physical degree of freedom
relates to the temperature
via $T\simeq 1.037\epsilon$, and
inserting this and the value of the reduced Planck 
mass $m_{Pl}=2.43\times 10^{18}$ GeV
in (\ref{eq2}) gives 
\begin{equation}\label{eq3}
m_s=\frac{m_{Pl}}{61.9} =3.9\times 10^{16}\,\mbox{GeV}.
\end{equation}

This simple calculation also shows that the string scale for 
type {\sl II} theories
and type {\sl I} theory or compactified 11D supergravity
would have to be higher to explain the same measured value
of the Planck scale. 
However, from the
string theory point of view the scale $m_s$ is 
more fundamental than $m_{Pl}$,
the latter being fixed through the string scale and the 
requirement that the
transition to the field theory description has to emerge self-consistently
at an energy density where string lengths can be resolved (\ref{eq2}).
Therefore, if we would not see
low energy remnants of
a particle spectrum inherited from heterotic string theory but 
rather detect
remnants of type {\sl II}, type {\sl I}, or SUGRA spectra, we 
would see
a smaller value of the Planck mass: Gravitational systems are 
stronger bound
in these phases.

The matching condition may also explain the 
dominance of the heterotic phase in the low energy regime:
Assuming that the energy density of the universe decreases from a 
high value above
the heterotic or type {\sl II} string scale, which field theory should we 
expect to emerge?
To me it seems reasonable to assume 
that field theory to take over which satisfies the condition 
\begin{equation}\label{eq4}
\varrho=\frac{\,\pi^2}{30}g_{\ast}(T_s)T_s^4
\end{equation}
 for transition to a field theory description first. However,
the theory which satisfies this condition
 for the highest possible energy density for given string length is the 
theory which
contains most relativistic degrees of freedom, i.e.\ the heterotic string.

In deriving the ratio (\ref{eq3}) between the reduced Planck mass and the 
string scale
I assumed that the field theory emerging at the string scale 
is four-dimensional from
the outset. If there is an intermediate ten-dimensional field theory phase 
below the string scale
and above a separate Kaluza--Klein scale $m_{KK}$ this scale will also 
show up
in the matching condition at the string scale.
The dependence of $m_{Pl}/m_s$ on $(m_s/m_{KK})^{(n-3)/2}$ is in fact the 
strongest impact of an intermediate $(n+1)$-dimensional field theory phase
on the string scale.
To elucidate this consider radiation 
dominated Friedmann cosmology in $D=n+1$ dimensions: 
The energy density $\varrho$
and the
scale factor ${\cal R}$ follow from
\[
\dot{\cal R}^2=\frac{2\kappa_D}{n(n-1)}{\cal R}^2\varrho,
\]
\[
\frac{d\varrho}{\varrho}=-(n+1)\frac{d{\cal R}}{\cal R},
\]
where $\kappa_D$ multiplies the energy momentum tensor in 
the $D$-dimensional
Einstein equations. These equations are readily integrated to yield
\begin{equation}\label{rhont}
\varrho=\frac{2n(n-1)}{(n+1)^2\kappa_D t^2}.
\end{equation}
The thermodynamic expressions follow from the corresponding
phase space integrals in $n$ spatial dimensions
\begin{equation}\label{rhonT}
\varrho=g_{\ast}\frac{n!\zeta(n+1)}{2^{n-1}\sqrt{\pi}^n\Gamma(n/2)}T^{n+1},
\end{equation}
where bosonic and fermionic degrees of freedom contribute according to
\[
g_{\ast}=g_B+\Big(1-\frac{1}{2^n}\Big)g_F.
\]
The difference between the average energy per relativistic
degree of freedom and spatial direction $\epsilon$ and the temperature 
\[
\epsilon=\frac{2^{n+1}-1}{2^{n+1}-2}\frac{\zeta(n+1)}{\zeta(n)}T
\]
does not have a strong impact, but is nevertheless taken into 
account.

The inverse compactification volume rescales $1/\kappa_D$ to the 
reduced
Planck mass via $1/\kappa_D=m_{Pl}^2 m_{KK}^{n-3}$, and we 
can not directly infer
the ratio between $m_s$ and $m_{Pl}$ without further information 
on $m_{KK}$. 
In ten dimensions the effective number of degrees of freedom 
of the heterotic string 
is $g_{\ast 10}=8056.125$ and
the resulting estimates on the string scale
with an intermediate ten-dimensional phase is 
\[
m_{s10}=3.9\times 10^{16}\,\mbox{GeV}\times (m_{KK}/m_{s10})^3.
\]

However, M-theory in its current
stage of evolution indicates that there is nothing special 
about 1-branes
in ten dimensions, and
on a field theoretic level
neither string theory nor M-theory require an 
intermediate ten-dimensional field
theory phase to occur. For this reason
and because compactification is really hard to achieve
in a purely field theoretic setting 
I consider a direct transition to a four-dimensional field theory
as the most interesting option, thus favoring a string scale (\ref{eq3}).
After all M-theory will have to tell us the number of 
dimensions to occur
in the transition to field theory.

How can it be that heterotic string thermodynamics gives such a  
low value for 
the string scale which was considered impossible before? 
Gross {\em et al.} 
and Ginsparg
had already pointed out that the string and the gauge coupling 
should relate to 
the ratio between the string scale and the Planck scale
according to $m_s\simeq gm_{Pl}$ \cite{GHMR2,pg},
and Kaplunovsky had addressed 
the corresponding threshold corrections to string unification \cite{vk}.

Inserting our result into this relation would give too
low a value for $g$. However, the indentification of $g$ with the ratio
 $m_s/m_{Pl}$ was derived under the assumption that low energy
gravitational and gauge interactions can alternatively be described in 
terms of string graviton
and vector
exchange, thus matching the string coupling to the Planck scale. 
In the present
paper, on the other hand, the Planck mass in the first place 
appears as a constant
of proportionality between the energy density $\varrho_c$ needed
to resolve a string and the string length squared. From the 
low energy point of view $\varrho_c$ is much higher
than $m_s^4$ because there are so many degrees of freedom in the 
heterotic phase
to absorb energy before the heat bath can resolve strings or 
other extended
objects. From the 
high energy point of view $\varrho_c$ is much higher
than $m_s^4$ because there are so many degrees of freedom 
in the heterotic phase
to transfer energy into field theory degrees of freedom.
Reinspection of the calculations of Gross {\em et al.} and Ginsparg
in the light of Eq.\ (\ref{eq3}) implies that in the M-theory phase
graviton exchange is stronger in comparison to vector exchange 
than in the field theory phase by a factor $m_{Pl}/m_s$, and this 
may shed new light on
the relation between M-theory interactions and low energy 
gauge and gravitational
interactions. This also indicates that the string scale is not just a scale
where a field theory approximation to M-theory becomes 
poor but rather corresponds
to a phase transition.

\end{document}